\pdfoutput=1

\documentclass[11pt, oneside]{article}   	
\usepackage{geometry}                		
\usepackage[parfill]{parskip}    		    
\usepackage{graphicx}				        

\usepackage{amssymb}
\usepackage{amsmath}

\usepackage{bm}
\usepackage{hyperref}

\usepackage{authblk}

\newcommand{\imp}{\mathcal{I}}

\newcommand{\vlrg}{V_{lrg}}
\newcommand{\vsml}{V_{sml}}
\newcommand{\vbz}{{V^b_0}}
\newcommand{\vaz}{{V^a_0}}

\newcommand{\pup}{p_\uparrow}
\newcommand{\pdn}{p_\downarrow}
\newcommand{\pex}{p_{ex}}
\newcommand{\puu}{p_{\uparrow\uparrow}}
\newcommand{\pud}{p_{\uparrow\downarrow}}
\newcommand{\pdu}{p_{\downarrow\uparrow}}
\newcommand{\pdd}{p_{\downarrow\downarrow}}

\newcommand{\hup}{\hat p_\uparrow}
\newcommand{\hdn}{\hat p_\downarrow}
\newcommand{\hex}{\hat p_{ex}}
\newcommand{\huu}{\hat p_{\uparrow\uparrow}}
\newcommand{\hud}{\hat p_{\uparrow\downarrow}}
\newcommand{\hdu}{\hat p_{\downarrow\uparrow}}
\newcommand{\hdd}{\hat p_{\downarrow\downarrow}}

\newcommand{\heu}{\hat p_{ex,\uparrow}}
\newcommand{\hed}{\hat p_{ex,\downarrow}}

\newcommand{\huz}{\hat p_{\uparrow0}}
\newcommand{\hdz}{\hat p_{\downarrow0}}
\newcommand{\hez}{\hat p_{ex,0}}

\newcommand{\bup}{\bar p_\uparrow}
\newcommand{\bdn}{\bar p_\downarrow}
\newcommand{\bex}{\bar p_{ex}}
\newcommand{\buu}{\bar p_{\uparrow\uparrow}}
\newcommand{\bud}{\bar p_{\uparrow\downarrow}}
\newcommand{\bdu}{\bar p_{\downarrow\uparrow}}
\newcommand{\bdd}{\bar p_{\downarrow\downarrow}}

\newcommand{\vecv}{\bm{V}}

\title{Apparent impact: the hidden cost of one-shot trades}
\author{Iacopo Mastromatteo}
\affil{\small Centre de Math\'ematiques Appliqu\'ees, CNRS, \'Ecole Polytechnique, \authorcr UMR 7641, 91128 Palaiseau, France}
\date{}							

\begin{document}
\maketitle

\begin{abstract}
We study the problem of the execution of a moderate size order in an illiquid market within the framework of a solvable Markovian model. We suppose that in order to avoid impact costs, a trader decides to execute her order through a unique trade, waiting for enough liquidity to accumulate at the best quote. We find that despite the absence of a proper price impact, such trader faces an execution cost arising from a non-vanishing correlation among volume at the best quotes and price changes. We characterize analytically the statistics of the execution time and its cost by mapping the problem to the simpler one of calculating a set of first-passage probabilities on a semi-infinite strip. We finally argue that price impact cannot be completely avoided by conditioning the execution of an order to a more favorable liquidity scenario.
\end{abstract}

\section{Introduction}
Market impact refers to the expected price change after the sequential execution of a given volume of contracts in a financial market~\cite{Bouchaud:2010aa}. It refers to one of the most fundamental aspects of market microstructure, as it encompasses the information about how a financial market reacts to an incoming fluxes of orders, and ultimately allows prices to reflect fundamental information~\cite{Bouchaud:2009aa}. Market impact is a central concept also for practitioners, who need to split their large orders (also called meta-orders) in sequences of smaller size child-orders in order to minimize transaction costs~\cite{bertsimas1998optimal}.

The existence of such \emph{order-splitting} procedures raises a practical question: Starting from what volume is it appropriate to split a meta-order, and when is it wiser to execute by using a single trade? Even though the instantly available liquidity is often insufficient in order to instantly execute a large meta-order at the best available price, for moderate size meta-orders it could be convenient to wait opportunistically until a sufficiently large fluctuation of the outstanding liquidity is realized, so to clear the meta-order at the best quote with a single trade.
While intuitively one would place the threshold of volume a few error bars away from the average value of the outstanding volume, it would be desirable to have a model precisely quantifying how to optimally perform this choice.

In this work we analyze such a \emph{one-shot} execution strategy within a specific solvable framework, identifying its average cost and characterizing its statistical fluctuations. Interestingly, we find that despite the absence of a proper market impact such a one-shot execution strategy has a positive average cost.
In particular, we find that whenever the outstanding liquidity is correlated with the direction of past price moves, waiting for sufficient volume to accumulate at the best quote can be expensive.

More importantly, we find that such a cost would be detected in data as an extra component of the \emph{slippage} (i.e., the expected price change between the \emph{decision time} and the \emph{execution time} of a meta-order) previously neglected in the literature, which is unrelated to the traditional notion of market impact and that is linked to the microstructural liquidity profile of the traded contracts. Yet, we dub it \emph{apparent impact}, as in empirical data it would appear as a component of the curve describing the transient impact of a meta-order emerging in absence of a price trend or a short-term alpha, and even in the case in which no trade at all is performed.

The reason why such a component of the impact should be properly taken into account in the estimation of the impact function is the extremely small entity of the price signal: while the fluctuations of daily returns are typically of the order of a few \%, the impact generally consists of some basis points, implying a signal to noise ratio around $10^{-2}$~\cite{Bouchaud:2009aa}. This forces one to carefully remove even very small sources of bias --~such as the aforementioned apparent component of impact~-- from the raw impact curves, in order not to compromise their statistical analysis. In fact, while the most prominent features of market impact (e.g., its concave dependence upon the executed volume) seem to be roughly universal~\cite{Torre:1997aa,Almgren:2005aa,Moro:2009aa,Toth:2011aa,Kyle:2012aa,Bershova:2013aa,Waelbroeck:2013aa,Mastromatteo:2014aa,bacry2014market,zarinelli2014beyond}, there is no widely-shared consensus about its detailed structure (e.g., the precise value of the exponent, the dependence of the impact function upon the trading strategy and the participation rate).

We will focus our attention on a specific scenario in which these ideas can be addressed analytically. In particular we will consider a stylized version of the framework presented in~\cite{Cont:2010aa,Cont:2013aa}, addressing the problem of a trader who needs to submit a buy meta-order of size $Q$ to the market (in an arbitrary amount of time), and who decides to wait until a sufficient amount of liquidity is available at the best quote before executing it by using a single trade, in order not to incur impact costs. We will be able to argue that:
\begin{itemize}
\item Between the \emph{decision time} and the \emph{execution time} the price will drift on average, even in absence of trend, by an amount $\imp(Q)$, where $\imp(\cdot)$ is a function of the executed volume, whose precise shape depends upon the details of market microstructure;
\item Such a price change is not linked to genuine impact (neither mechanical or behavioral), as we take as the execution time the instant \emph{right before} the submission of the order, in which the trader has not yet executed her order.
\end{itemize}
Our main conclusion is that past price changes can induce future trading activity simply for liquidity reasons: If less liquidity is available at the ask whenever prices have a bullish trend, then buyers need to be more patient while price moves up. Conversely, they can execute more quickly when prices decline. We believe that, irrespective of the particular model that we have chosen in order to illustrate this behavior, such phenomenon arises with greater generality, and that it generates additional execution costs with respect to the traditional market impact. Our framework should then be regarded as a minimal model capturing the emergence of the apparent component of the impact function, allowing to explore analytically the relation among liquidity at the best quote, local price trend and speed of execution.

The plan of the paper is the following: In Sec.~\ref{sec:model} we introduce our market model, reviewing its properties and introducing the notation adopted in the following sections. The main results are given in Sec.~\ref{sec:one_shot}, where we present the order execution procedure for the extra trader and calculate the expected properties of the model at the moment of the execution. Sec.~\ref{sec:discussion_and_extensions} presents a critical assessment of the modeling assumptions, and discusses several generalizations. Finally, we draw our conclusions in Sec.~\ref{sec:concl}. For the sake of clarity, the more technical parts are relegated to the appendices. 

\section{The model}
\label{sec:model}
\subsection{A Markovian description of large tick contracts}
We adopt a framework inspired by the one considered in~\cite{Cont:2010aa,Cont:2013aa} as a background model of the market. In particular, we consider a Markovian model in which all the information relative to the current state of the market is encoded in these variables:
\begin{itemize}
\item The mid-price $x_t$;
\item The volume available at the bid queue $V^b_t$ and the volume available at the ask queue $V^a_t$.
\end{itemize}
This choice is motivated by the strong concentration of market activity at the best quotes for large tick stocks~\cite{Biais:1995aa}, and by the observation that a large component of price variations  can be accounted for by the dynamics at the best quotes~\cite{Eisler:2012aa,Cont:2013ab}. In order to model the mechanism of quote revision, we suppose that anytime either queue is exhausted, the volume of both queues is reset and prices are updated accordingly. In particular if the ask queue is emptied, we assume that the volumes at the best queues $\vecv=(V^b,V^a)$ instantaneously revert to $(\vsml,\vlrg)$, while in the opposite case we assume the new volumes to be $(\vlrg,\vsml)$, which allows to model microscopic mean-reversion of price (see e.g.~\cite{Cont:2013ab}). We suppose in the former case the price to jump up by one half of a tick (which we assume without loss of generality to be of size $w=2$), while in the latter case we assume price to decrease by $w/2$. This should be appropriate in the case of large tick contracts, for which the spread is closed almost immediately after the exhaustion of the volume at either of the best queues~\cite{Gareche:2013aa}.

For the dynamics of the queues we choose a stylized version of the one first considered in~\cite{Cont:2010aa}, in which we disregard the granularity in the dynamics (the Poissonian nature of the volume jumps) and focus on the diffusive limit derived in~\cite{Cont:2013aa} (see Sec.~\ref{sec:discussion_and_extensions} for a critical assessment of our assumptions, and a list of possible generalizations of our approach).
Within our simplified description of the queues dynamics, the volumes at the best quotes evolve according to the equation
\begin{equation}
\label{eq:diff_queue}
  \frac{\partial P(t,V^b,V^a)}{\partial t} = \frac{D}{2} \frac{\partial^2 P(t,V^b,V^a)}{\partial (V^b)^2} +\frac{D}{2} \frac{\partial^2 P(t,V^b,V^a)}{\partial (V^a)^2} 
  + \mu  \frac{\partial P(t,V^b,V^a)}{\partial V^b}  + \mu  \frac{\partial P(t,V^b,V^a)}{\partial V^b}  \; ,
\end{equation}
where we take $D=1$ without loss of generality and where we assume $\mu \geq0$, so to model queues whose volume drifts towards zero.
Under this dynamics no detailed description of the distinct type of orders hitting the best queues is provided\footnote{Notice that in this framework market orders and cancellations are not distinguishable.}. Rather, this scenario builds on a coarse-grained description of the volume available at the best quotes which is appropriate when volumes sitting at the best queues are sufficiently large.

Finally, the procedure of one-shot execution of a buy order illustrated above implies in this context in the presence of a random variable $T$ representing the time elapsed between the \emph{decision time} $t=0$ and the \emph{execution time} $t=T$, the first time in which the ask volume $V^a$ equals $Q$. Our goal is to provide a statistical characterization of $T$, together with the one of the price $x_t$ (the number of positive jumps minus the negative ones) and the one for hitting number $n_t$ (the number of jumps of either type) at the execution time $T$.

\subsection{Properties of the free model}
\label{sec:free}
At first we will illustrate the properties of this model in the regime in which no trader is present, and the two queues evolve freely under the dynamics (\ref{eq:diff_queue}) and the boundary conditions specified above. We use the word \emph{freely} in order to indicate the unconditional evolution of the queues (i.e., the absence of the extra-trader), as opposed the one in which volumes at the ask cannot do not exceed the value of $Q$ (i.e., when the extra-trader is present). In the free case, the coordinates of the system $(V^b,V^a)$ can then diffuse in the whole positive orthant, reverting to the appropriate state (either $(\vlrg,\vsml)$ or $(\vsml,\vlrg)$) as soon as they touch either of the boundaries. In either case, the variables $x_t$ and $n_t$ are conveniently updated. We are interested in particular in showing the results for the statistics of the evolution of the price $x_t$ and the hitting number $n_t$. The  object that we are required to compute in order to obtain a solution for $x_t$ and $n_t$ is the probability that a system starting from coordinates $(V^b,V^a)$ hits one of the the two boundaries (either $V^a=0$ or $V^b=0$) at a given time $t$. These quantities will be denoted respectively with $\pup(t,V^b,V^a)$ and $\pdn(t,V^b,V^a)$. It is easy to show (see for instance Ref.~\cite{Kearney:2005aa}) that their Laplace transforms $\hup(\omega,V^b,V^a)$ and $\hdn(\omega,V^b,V^a)$ satisfy the equations
\begin{eqnarray}
\label{eq:hit_laplace}
\omega \, \hat p_\alpha(\omega,V^b,V^a) &=& \frac{1}{2} \frac{\partial^2 \hat p_\alpha(\omega,V^b,V^a)}{\partial (V^b)^2} + \frac{1}{2}\frac{\partial^2 \hat p_\alpha(\omega,V^b,V^a)}{\partial (V^a)^2}  \nonumber \\
&-& \mu \frac{\partial \hat p_\alpha(\omega,V^b,V^a)}{\partial V^b} -\mu \frac{\partial \hat p_\alpha(\omega,V^b,V^a)}{\partial V^a}\; ,
\end{eqnarray}
where $\alpha \in \{ \uparrow, \downarrow \}$. The difference among $\pup$ and $\pdn$ is encoded in the different boundary conditions: for  $V^b\neq 0$ one has $\pup(t,V^b,0) = \delta_t$,  $\pdn(t,V^b,0) = 0$, while for $V^a\neq0$ one has $\pdn(t,0,V^a) = \delta_t $, $\pup(t,0,V^a) =0$. Equivalently $\hup(\omega,V^b,0)=1-\hdn(\omega,V^b,0) = 1$,  $\hdn(\omega,0,V^a)=1-\hup(\omega,0,V^a) = 1$.
The values $\bup(V^b,V^a)= \hup(0,V^b,V^a)$ and $\bdn(V^b,V^a)=\hdn(0,V^b,V^a)$ obtained for $\omega=0$ represent the probability of hitting the boundaries at any time between $t=0$ and $t=\infty$ starting from the configuration $(V^b,V^a)$, and verify $\bup(V^b,V^a) +\bdn(V^b,V^a) = 1$, indicating that the boundaries are hit almost surely.
Once that these functions have been calculated (the details about the derivation can be found in App.~\ref{app:prob_abs}), it is possible to obtain the generating function for the price $x_t$, denoted with $ \Phi_x(\omega,s)$, as well as the one for the hitting number $n_t$, denoted with $ \Phi_n(\omega,s)$. They are defined as 
\begin{eqnarray}
\label{eq:free_gf_price}
\Phi_x(\omega,s)&=& \sum_{x=-\infty}^{\infty} e^{-xs} \int_0^\infty dt\, e^{-\omega t}  P_x(t,x)\\
\label{eq:free_gf_hit}
\Phi_n(\omega,s)&=& \sum_{n=-\infty}^{\infty} e^{-ns} \int_0^\infty dt\, e^{-\omega t} P_n(t,n) \; ,
\end{eqnarray}
where $P_x(t,x)$ and $P_n(t,x)$ are respectively the probability for the price and for the hitting number of taking value $x$ (respectively $n$) at time $t$. By exploiting the Markov property of the model, one can prove (App.~\ref{app:markov_chain}) that
\begin{eqnarray}
\label{eq:free_markov_price}
  \Phi_x(\omega,s) &=& \frac{1}{\omega} \left[  
    \left(
    \begin{array}{c}
    1-\huu-\hdu \\
    1-\hud-\hdd
  \end{array}
  \right)^T
  \sum_{n=0}^\infty
\left(
  \begin{array}{cc}
    \huu e^{-s}& \hud e^{-s} \\
    \hdu e^{s} & \hdd e^s
  \end{array}
\right)^n
\left(
  \begin{array}{c}
    \huz e^{-s} \\
    \hdz e^s
  \end{array}
\right)
\right] \\
&+&
\frac 1 \omega \left(  1 -\huz - \hdz \right)
\; , \nonumber  \\
\label{eq:free_markov_hit}
  \Phi_n(\omega,s) &=& \frac{1}{\omega} \left[
      \left(
    \begin{array}{c}
    1-\huu-\hdu \\
    1-\hud-\hdd
  \end{array}
  \right)^T
  \sum_{n=0}^{\infty}
\left(
  \begin{array}{cc}
    \huu & \hud \\
    \hdu & \hdd
  \end{array}
\right)^n
\left(
  \begin{array}{c}
    \huz \\
    \hdz
  \end{array}
\right)
e^{-(n+1)s}
\right] \\
&+&
\frac 1 \omega \left(  1 -\huz - \hdz \right)
\; , \nonumber 
\end{eqnarray}
where we have introduced the notation
\begin{eqnarray}
\huz = \hup(\omega,\vbz,\vaz)&& \hdz = \hdn(\omega,\vbz,\vaz) \\
\huu = \hup(\omega,\vsml,\vlrg)&& \hdu = \hdn(\omega,\vsml,\vlrg) \\
\hud = \hup(\omega,\vlrg,\vsml)&& \hdd = \hdn(\omega,\vlrg,\vsml) \, ,
\end{eqnarray}
and $(\vbz,\vaz)$ are the coordinates from which the system starts at $t=0$. Equations~(\ref{eq:free_markov_price}) and~(\ref{eq:free_markov_hit}) are readily solved by diagonalization of their respective transition matrices, as shown in App.~\ref{app:markov_chain}.
In the free case the \emph{bid-ask symmetry} holds (i.e,  $\pup=\pdn$, $\puu=\pdd$ and $\pud=\pdu$). We will show in Sec.~\ref{sec:one_shot} that the presence of a extra agent passively waiting for a volume $Q$ breaks this symmetry, leading to the emergence of the aforementioned apparent component of the impact.

As the the behavior of the model changes qualitatively in the cases $\mu=0$ and $\mu > 0$, we will discuss the phenomenology of the model in the two cases separately.

\subsection{The driftless case}
\begin{figure}[p]
  \centering
  \includegraphics{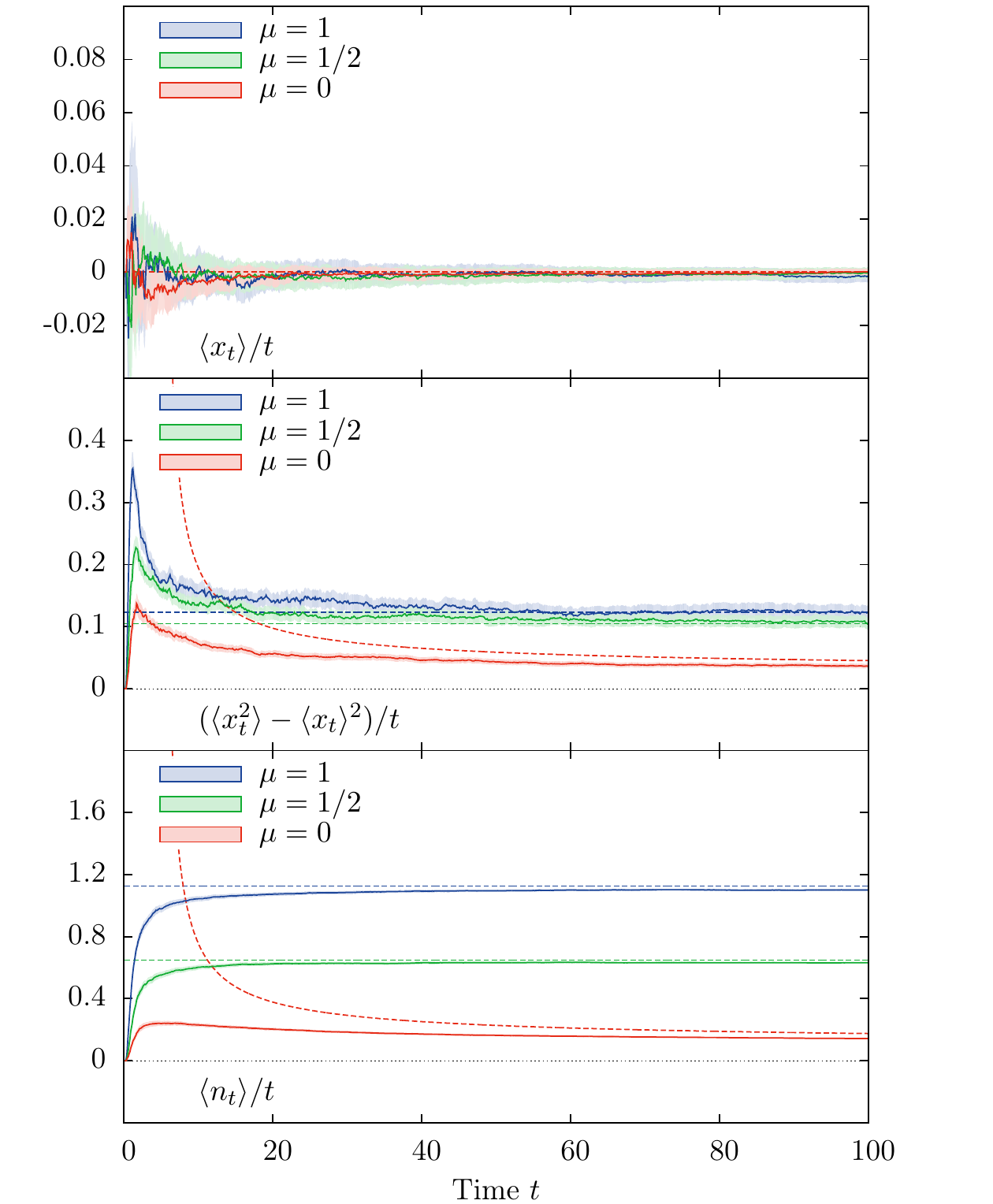}
  \caption{Evolution in time of the average price change $\langle x_t \rangle$, hitting number $\langle n_t \rangle$ and variogram $(\langle x_t^2\rangle - \langle x_t \rangle^2)/t$ for the free diffusion problem. The solid lines indicate the result of simulations for which we used the set of parameters $V_0=2,\vsml=1,\vlrg=3$, while the shaded regions account for two-sigma statistical fluctuations around the mean values obtained by averaging over 1000 realizations. The different curves describe both the driftless case ($\mu=0$, corresponding to the red line) and the drifted case ($\mu=1/2$ and $\mu=1$ are associated respectively with the green and the blue line). The dashed lines indicate the asymptotic predictions displayed in Sec.~\ref{sec:model}.}
  \label{fig:free_diff}
\end{figure}

In the driftless case the volumes at the best almost surely drop to zero for any starting condition (one can in fact verify that the sum $\bup + \bdn$ is equal to one). The exhaustion of the volume can indeed take a very large time, as the volume fluctuations are anomalously large. This leads to a logarithmic sub-diffusion of the price in the long time limit. The assumption $\vsml < \vlrg$ induces short-time mean reversion --~in accordance with well-established empirical evidence concerning the short-time behavior of the volatility function~-- partially damping the asymptotic value of the price fluctuations. We will illustrate this behavior in the following section.\\
The absorption probabilities in the driftless scenario can be calculated explicitly (App.~\ref{app:prob_abs}), and their values may be expressed as the integrals
\begin{eqnarray}
  \label{eq:free_abs_dless_up}
  \hup (\omega,V^b,V^a) &=& \frac{2}{\pi} \int_0^\infty dk \frac{\sin(kV^a)}{k} \frac{1}{1+2\omega /k^2} (1-e^{-\sqrt{k^2 +2\omega}\;V^b})  \\
  \label{eq:free_abs_dless_dn}
  \hdn (\omega,V^b,V^a) &=& \frac{2}{\pi} \int_0^\infty dk\frac{\sin(k V^a)}{k} e^{-\sqrt{k^2 +2\omega}\;V^b} \; .
\end{eqnarray}
The values for $\omega=0$, expressing the probability of the sign of the next price change, are more conveniently written in terms of elementary functions, and result
\begin{eqnarray}
  \label{eq:26}
  \bup (V^b,V^a) &=&  \frac{2}{\pi} \arctan\left(\frac{V^b}{V^a}\right) \\
  \bdn (V^b,V^a) &=&  \frac{2}{\pi} \arctan\left(\frac{V^a}{V^b}\right) \; .
\end{eqnarray}
The mean absorption time by either of the boundary is infinite (equivalently, the derivatives of Eqs.~(\ref{eq:free_abs_dless_up}) and~(\ref{eq:free_abs_dless_dn}) with respect to $\omega$ are divergent at $\omega=0$).
The explicit expression of the absorption probability then needs to be inserted into the generating functions~(\ref{eq:free_gf_price_app}) and~(\ref{eq:free_gf_hit_app}) in order to estimate numerically the momenta of the price change and the hitting times. A small $\omega$ expansion (performed in App.~\ref{app:prob_abs}) allows to evaluate the large time behavior of these quantities, which results\\
\begin{eqnarray}
    \langle x_t \rangle &=& 0 \\
  \label{eq:dless_vol}
  \langle x_t^2 \rangle -  \langle x_t \rangle^2 &=&  \frac{ \chi\, t}{\log t} \left( \frac \pi {2 \vsml \vlrg} \right) + o(t/\log t) \\
  \langle n_t \rangle &=& \frac{t}{\log t} \left( \frac \pi {2 \vsml \vlrg} \right) + o(t/\log t) \\
  \langle n^2_t \rangle -  \langle n_t \rangle^2 &=& \frac{t^2}{\log^3(t)}\left(\frac{\pi}{2\vsml\vlrg} \right)^2 + o(t^2/\log^3t) \; ,
\end{eqnarray}
where we have introduced the asymmetry parameter $\chi=\huu/\hdu=\hdd/\hud$ accounting for the short-time mean-reversion of price.
An inspection Eq.~(\ref{eq:dless_vol}) reveals that the behavior of price is sub-diffusive, as the variance of price increases less then linearly. This behavior derives from the broad tails of the distribution for the absorption time, whose anomalously large fluctuations cause the mean absorption time to diverge.  Fig.~\ref{fig:free_diff} summarizes these results by showing the evolution in time of the price change $x_t$ and the hitting number $n_t$, and by comparing it with the results of numerical simulations.
\subsection{The case $\mu > 0$}
The phenomenology of a system in which the drift $\mu$ is non-zero is to some extent similar to the one discussed above, as the volumes at the best fall to zero almost surely. The most notable difference is that the average absorption time by either of the boundaries is finite. This leads to a diffusive behavior of the price at large times, while at short times the price has a mean-reverting behavior for $\vsml < \vlrg$ similar to the one observed in the driftless case.

The Laplace transforms of the absorption probabilities for $\mu>0$ can be expressed as the integrals
\begin{eqnarray}
  \label{eq:free_abs_drift_up}
  \hup (\omega,V^b,V^a) &=& \frac{2}{\pi} \int_0^\infty dk \frac{\sin(kV^a)}{k} \frac{e^{\mu V^a}}{1+(2\omega +\mu^2)/k^2} \left[1-e^{-\left(\sqrt{k^2 +2\omega+2\mu^2}-\mu\right) \;V^b}\right] \\
  \label{eq:free_abs_drift_dn}
  \hdn (\omega,V^b,V^a) &=& \frac{2}{\pi} \int_0^\infty dk\frac{\sin(k V^a)}{k}  \frac{e^{\mu V^a}}{1+\mu^2/k^2} e^{-\left(\sqrt{k^2 +2\omega+2\mu^2}-\mu\right) \;V^b} \; .
\end{eqnarray}
while the  probabilities for the sign of the next price change are recovered by setting $\omega=0$ in above formula. \\
The average absorption times can be obtained by differentiation of Eqs.~(\ref{eq:free_abs_drift_up}) and~(\ref{eq:free_abs_drift_dn}), which lead to convergent integrals, as opposed to the driftless scenario discussed above in which the resulting expressions were divergent.
\\
These results have been used in Fig.~\ref{fig:free_diff} in order to show the large time asymptotics for the mean values of $x_t$ and $n_t$ and their fluctuations  in the case $\mu\neq0$. In fact the large time behavior can be written explicitly by expanding the generating functions close to the point $\omega=0$. We find at leading order in time:
\begin{eqnarray}
    \langle x_t \rangle &=& 0 \\
  \langle x_t^2 \rangle -\langle x_t \rangle^2 &=&  \frac{ t}{\langle t_{hit}\rangle}\,\chi + O(1) \\
  \langle n_t \rangle &=& \frac{t}{\langle t_{hit}\rangle} + O(1) \\
  \langle n^2_t \rangle  -\langle n_t \rangle^2&=& \left( \frac{\langle t_{hit}^2 \rangle-\langle t_{hit} \rangle^2}{\langle t_{hit} \rangle^3} \right) t + O(1) \; ,
\end{eqnarray}
where for $q\in\mathbb{N}$ we have defined the momenta $\langle t_{hit}^q\rangle=\langle t_{hit,\uparrow}^q\rangle=\langle t_{hit,\downarrow}^q\rangle$ relative to the average time required to hit any of the boundaries starting from either initial condition.
The above expression allows to relate the (asymptotic) \emph{volatility} $\sigma^2=\lim_{t\to\infty}\langle x_t^2\rangle/t = \chi /\langle t_{hit} \rangle$ to the microstructural parameters governing the model.

\section{One-shot execution}
\label{sec:one_shot}
In order to illustrate the notion of apparent impact in our scenario, we consider a setting in which the trader starts waiting for a volume $Q$ to accumulate on the ask queue at an initial time $t=0$ in which the system is characterized by coordinates $(\vbz,\vaz)$, with $\vbz<Q$. As we are interested in characterizing the statistics of the price change $x_t$ and the hitting number $n_t$ as soon the volume at the ask queue reaches a volume $Q$, we define a stopping time $T$ at which $V^a=Q$ at which the trader executes her order with a single trade.

The main change with respect to the free case discussed above is the fact that the diffusion of the coordinates $(V^b,V^a)$ no longer takes place on the positive orthant $(V^b,V^a)\in(0,\infty) \times (0,\infty)$, but on the semi-infinite strip $(V^b,V^a)\in(0,\infty)\times(0,Q)$. We need correspondingly to define a third type of probability associated with the absorption by the $V^a=Q$ boundary, which we denote as $\pex(t,V^b,V^a)$. Its Laplace transform evolves according to equation~(\ref{eq:hit_laplace}) with the boundary conditions $\hex(\omega,V^b,Q)=1-\hup(\omega,V^b,Q) =1-\hdn(\omega,V^b,Q) = 1$. Also in this case it is possible to show that $\bup(V^b,V^a)+\bdn(V^b,V^a)+\bex(V^b,V^a) = 1$.
We remark that the bid-ask symmetry present in the free case is broken by the presence of the $V^a=Q$ boundary, so that an asymmetric evolution of the price is expected in this setting. This is, in a nutshell, the reason why the apparent component of the impact emerges in our model.

One can define the generating functions for the price $x_t$ and the hitting number $n_t$ which are relevant for this problem by supposing that, as soon as the boundary $V^a=Q$ is reached, the price and the hitting number are frozen at their respective values at time $t=T$. We denote the relevant probabilities for this problem as $P^{ex}_x(T,x)$ and $P^{ex}_n(T,x)$, respectively the probability for the price and the hitting number to assume the value $x$ and $n$ when at the stopping time $T$, and define their corresponding generating functions as
\begin{eqnarray}
\label{eq:gf_price_abs}
 \Psi_x(\omega,s)&=& \sum_{x=-\infty}^{\infty} e^{-xs} \int_0^\infty dT\, e^{-\omega T}  P^{ex}_x(T,x)\\
\label{eq:gf_hit_abs}
\Psi_n(\omega,s)&=& \sum_{n=-\infty}^{\infty} e^{-ns} \int_0^\infty dT\, e^{-\omega T} P^{ex}_n(T,n) \; .
\end{eqnarray}
These functions satisfy the set of equations
\begin{eqnarray}
\label{eq:gf_markov_price}
 \Psi_x(\omega,s) &=&   
  \hez +
    \left(
    \begin{array}{c}
    \heu \\
    \hed
  \end{array}
  \right)^T
  \sum_{n=0}^\infty
\left(
  \begin{array}{cc}
    \huu e^{-s}& \hud e^{-s} \\
    \hdu e^{s} & \hdd e^s
  \end{array}
\right)^n
\left(
  \begin{array}{c}
    \huz e^{-s} \\
    \hdz e^s
  \end{array}
\right)
 \\
\label{eq:gf_markov_hit}
 \Psi_n(\omega,s) &=&
  \hez +
      \left(
    \begin{array}{c}
   \heu \\
   \hed
  \end{array}
  \right)^T
  \sum_{n=0}^{\infty}
\left(
  \begin{array}{cc}
    \huu & \hud \\
    \hdu & \hdd
  \end{array}
\right)^n
\left(
  \begin{array}{c}
    \huz \\
    \hdz
  \end{array}
\right)
e^{-(n+1)s}
\, ,
\end{eqnarray} 
and as in the previous case can be solved by diagonalization of their respective transition matrices (App.~\ref{app:markov_chain}).

Summarizing, in order to characterize analytically the price $x_T$ and the hitting number $n_T$ at the moment preceding the trade, we need to solve the diffusion problem~(\ref{eq:diff_queue}) in the modified geometry of a semi-infinite strip, and successively plug the probabilities of hitting the boundaries into the generating functions $\Psi_x(\omega,s)$ and $\Psi_n(\omega,s)$, whose explicit expressions have been worked out in App.~\ref{app:markov_chain}.

\subsection{Expected slippage and average hitting number}
The qualitative behavior of the price change $x_t$ at the time of execution $t=T$ is similar in the $\mu=0$ and the $\mu>0$ scenarios: In both cases the apparent impact, that we identify with the average price change $\imp(Q) = \langle x_T \rangle$ increases --~in general non-monotonically~-- with the executed volume $Q$ from the initial value $\langle x_T \rangle=0$ corresponding to $Q=V_0$ to an asymptotically constant value reached for $Q\to\infty$. The hitting number $\langle n_T\rangle$ always increases monotonically in $Q$, growing unbounded from the value $\langle n_T \rangle=0$ at $Q=V_0$ up to infinity. The precise form of the asymptotic scaling in $Q$ depends crucially on $\mu$: while for $\mu=0$ the growth of $\langle n_T \rangle$ is asymptotically quadratic in $Q$, for $\mu> 0$ the growth is exponential, signaling that if the volumes at the best queues are drifted towards zero, one needs to wait an exponentially long number of price changes in order to execute a large order.\\
A similar behavior is observed for the price fluctuations $\langle x_T\rangle^2-\langle x_T\rangle^2$, implying that a patient trader who is willing to execute a large order needs to be ready to face a potentially large volatility risk.
Notice that at large $T$ one has $\langle n_T \rangle \sim (\langle x_T\rangle^2-\langle x_T\rangle^2)^{1/2} \gg \langle x_T\rangle$, due to the fact that the asymmetry induced in the diffusion problem by the absorbing boundary $V^a=Q$ is a smaller order effect with respect to the dominating symmetric behavior associated with the $Q\to\infty$ limit. 
Practically, this implies that the average value of the apparent impact is at most of the order of one tick, while the variance of price is expected to grow linearly in time as in the free case.
We also remark that the evolution of the momenta of $x_t$ and $n_t$ display a point of non-analyticity for $Q=\vlrg$. This is explained with the change of regime of the hitting probabilities for $\vlrg =Q$: we have in fact assumed that $\bup=\bdn=1-\bex=0$ independently of $Q$ as long as $Q<\vlrg$ (in such case the trader can in fact execute all the volume instantly after a positive price change), while for $Q>\vlrg$ we assume the hitting probabilities to satisfy Eq.~(\ref{eq:hit_laplace}).  

\begin{figure}[p]
  \centering
  \includegraphics{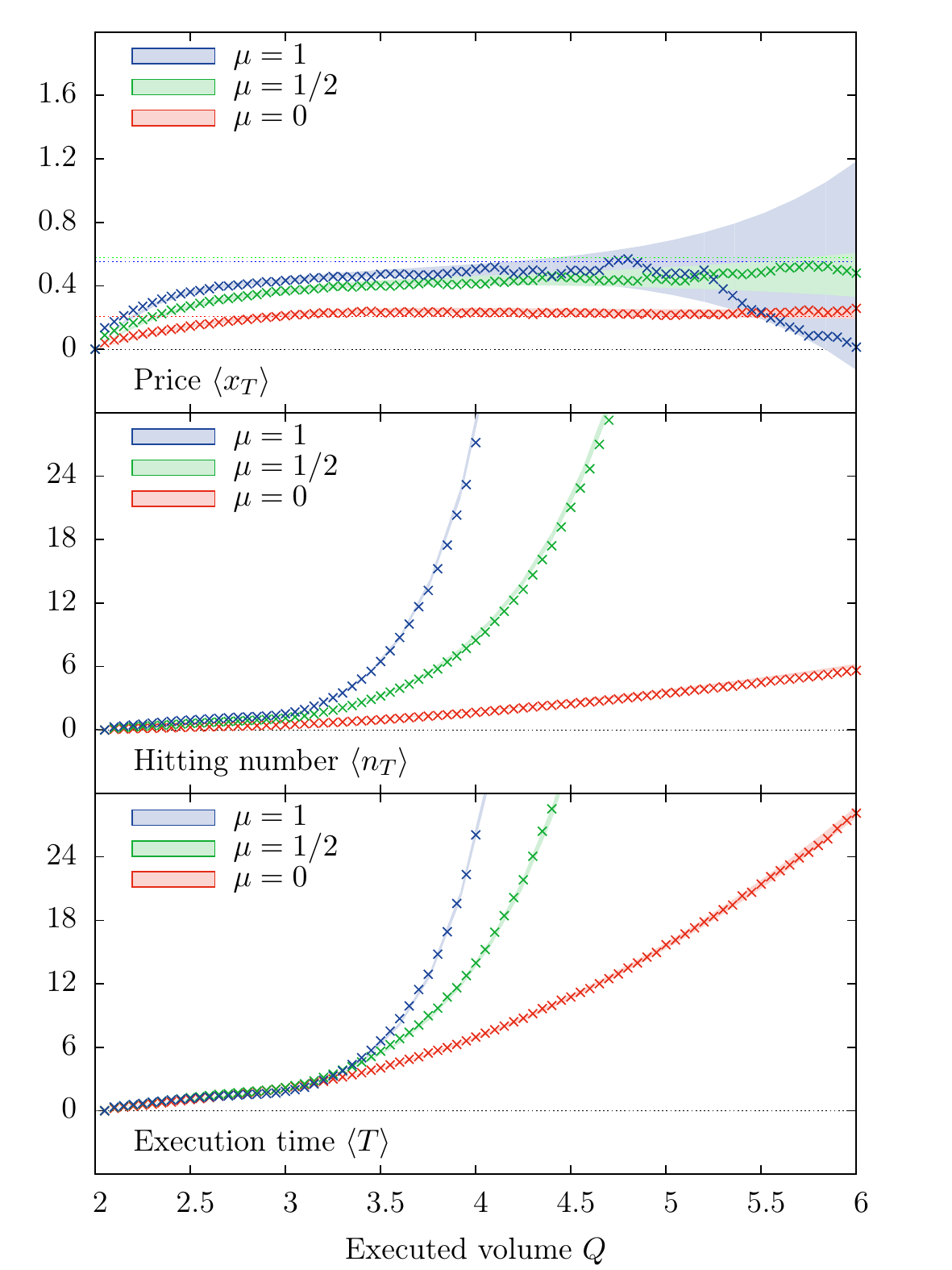}
  \caption{Statistics for the one-shot execution problem. We represent the averages of the price change $\langle x_t\rangle$,  hitting number $\langle n_t\rangle$ and execution time $\langle T \rangle$ as functions of the executed volume $Q$, for different values of the drift $\mu$. We have simulated 8000 realization of a process defined by the set of parameters $V_0=2,\vsml=1$ and $\vlrg=3$. The shaded regions correspond to the theoretical predictions, accounting for two-sigma regions, while the crosses are the results of numerical simulations. }
  \label{fig:exec}
\end{figure}

Fig.~\ref{fig:exec} show the result of a numerical simulation of the execution process, comparing it with the semi-analytical results obtained by integrating numerically the expressions for the average slippage and the hitting number.

The next part of this section will be devoted to a more detailed description the qualitative behavior sketched above. In particular we will use the results proved in the appendices in order to extract analytically the leading behavior of $x_T$ and $n_T$ at small and large $Q$.

By exploiting the results of the $1/Q$ expansion showed in App.~\ref{app:prob_abs}, we find in fact that for large executed volumes the apparent impact function tends to
\begin{equation}
  \label{eq:lrg_Q_av_price}
  \imp(Q) = \langle x_T \rangle  \xrightarrow[Q\to\infty]{} \left\{
  \begin{array}{lcr}
   \displaystyle \frac{1+\chi}{6} & \textrm{for}& \mu=0 \\[10pt]
   \displaystyle \frac{1+\chi}{2} & \textrm{for}& \mu> 0
  \end{array}
   \right.\; ,
\end{equation}
which indicates that for a market with short-time mean reversion of price ($\chi < 1$) the mean price change is asymptotically smaller than $\langle x_T \rangle=1/3$ (for $\mu=0$) or $\langle x_T \rangle=1$ (for $\mu >  0$).
The asymptotics of the average hitting number can also be extracted by means of a $1/Q$ expansion, and reads
\begin{equation}
  \label{eq:lrg_Q_av_hit}
    \langle n_T \rangle  \xrightarrow[Q\to\infty]{} \left\{
  \begin{array}{lcr}
   \displaystyle \frac{2Q^2}{\pi \vsml \vlrg} & \textrm{for}& \mu=0 \\[10pt]
   \displaystyle 2\sqrt{\pi } \mu^2 \left( \frac Q {2\sqrt{2} \mu }\right)^{3/2} \frac {e^{\mu  ((1+\sqrt{2})Q-\vsml-\vlrg)}} {\vsml \sinh \left(\sqrt{2} \mu  \vlrg\right)+\vlrg \sinh \left(\sqrt{2} \mu  \vsml\right) }  & \textrm{for}& \mu> 0
  \end{array}
   \right.\; ,
\end{equation}
so that the total number of price changes asymptotically grows either as $Q^2$ (in the case $\mu=0$) or with the exponential law $Q^{3/2}\exp(\mu(1+\sqrt{2})Q)$ (when $\mu>0$).
Finally, the price fluctuations obey the expected asymptotic scaling
\begin{equation}
  \label{eq:lrg_Q_flct_price}
    \langle x_T^2\rangle - \langle x_T^2 \rangle \xrightarrow[Q\to\infty]{} \chi \langle n_T \rangle \; ,
\end{equation}
which indicates that at leading order the asymptotic volatility of the free problem is not affected by the conditioning effect induced by the execution.

The small $Q$ asymptotics of the process (provided $\vsml < V_0 < \vlrg$) leads to a linear behavior in the vicinity of the point $Q=V_0$ for all the observables. In particular can use the relation $\huu = \hdu=0$ valid for $Q<\vlrg$ in order to show that
\begin{eqnarray}
  \label{eq:sml_Q_av_price}
  \langle x_T \rangle &\xrightarrow[Q\to V_0]{}& (Q-V_0)\left( \delta_\uparrow - \delta_\downarrow \frac {1-\hud}{1-\hdd} \right) \\
  \label{eq:sml_Q_av_hit}
    \langle n_T \rangle &\xrightarrow[Q\to V_0]{}& (Q-V_0)\left( \delta_\uparrow + \delta_\downarrow \frac {1+\hud}{1-\hdd} \right)\\
  \label{eq:sml_Q_flct_price}
    \langle x_T^2\rangle - \langle x_T^2 \rangle &\xrightarrow[Q\to V_0]{}& (Q-V_0)\left( \delta_\uparrow + \delta_\downarrow \frac{(1+\hdd)(1-\hud)}{(1-\hdd)^2} \right) \; ,
\end{eqnarray}
where we have defined
\begin{eqnarray}
  \label{eq:sml_Q_corr}
  \delta_\uparrow &=&\left\{
  \begin{array}{lcr}
    \displaystyle \frac 1 {V_0} \left( \frac{e^{\pi}-1}{e^{\pi}+1} \right) &\textrm{for}& \mu=0 \\[8pt]
    \displaystyle \frac 2 \pi \int_0^\infty \, dp \frac{\beta(p) p \sin(p V_0) }{p^2+\mu^2} \left( \frac{e^{2\mu V_0}}{\sinh(\beta(p)V_0)}\right)  &\textrm{for}& \mu>0
  \end{array}
\right.
\\
  \delta_\downarrow &=& \left\{
  \begin{array}{lcr}
    \displaystyle \frac 1 {V_0} \left(\frac 2 {\sinh \pi}\right) &\textrm{for}& \mu=0 \\[8pt]
    \displaystyle \mu + \frac 2 \pi \int_0^\infty \, dp \frac{\beta(p) p\sin(p V_0)  }{p^2+\mu^2} \left(\frac{\cosh(\beta (p)V_0) -e^{\mu V_0}}{\sinh(\beta(p)V_0)}\right) e^{\mu V_0} &\textrm{for}& \mu>0
  \end{array}
\right. \; ,
\end{eqnarray}
with $\beta(p)=\sqrt{p^2 + 2\mu^2}$. Hence, the small volume behavior of the apparent impact is analytical close to the starting volume $V_0$, in contrast with the ordinary price impact, which for small executed volumes  grows roughly as $Q^{\eta}$, being $\eta$ an exponent between 0.6 and 0.8 \cite{Torre:1997aa,Almgren:2005aa,Moro:2009aa,Toth:2011aa,Kyle:2012aa,Bershova:2013aa,Waelbroeck:2013aa,Mastromatteo:2014aa}.

\subsection{Average execution time}
The execution time $T$ follows a statistics similar to the one of the hitting number $n_T$, as showed in Fig.~\ref{fig:exec} where we display the evolution of the execution time as a function of the executed volume $V$. As for the average price change and the hitting number, we have compared with the simulations the semi-analytical result obtained by integrating numerically Eqs.~(\ref{eq:int_rep_hex}),~(\ref{eq:int_rep_hup}) and~(\ref{eq:int_rep_hdn}) for the hitting probabilities and inserting them into the Eqs.~(\ref{eq:abs_av_time}) and~(\ref{eq:abs_var_time}) for the momenta of $T$. We find that the average execution time increases monotonically from $\langle T \rangle=0$ for $Q=V_0$ up to an asymptotic regime whose scaling depends on the value of $\mu$.

In particular by performing a $1/Q$ expansion of the execution time, we find the asymptotic scaling
\begin{equation}
  \langle T \rangle \xrightarrow[Q\to\infty]{} \langle n_T \rangle \langle t_{hit} \rangle \; ,
\end{equation}
where we have used  the results of App.~\ref{app:tot_hit_time} in order to obtain the scaling of the hitting times
\begin{equation}
  \langle t_{hit,\alpha}\rangle \xrightarrow[Q\to\infty]{}
  \left\{
  \begin{array}{lcr}
    \displaystyle \frac {4 V^bV^a}{\pi} \log(Q)  &\textrm{for}& \mu=0 \\[8pt]
    \displaystyle \frac {V^b} \mu - \frac{4}{\pi} \int_0^\infty dp \frac{p\sin(pV^b)}{(p^2+\mu^2)^2} e^{-(\sqrt{p^2 + 2\mu^2} -\mu) V^a+\mu V^b} &\textrm{for}& \mu>0 
  \end{array}
  \right. \; ,
\end{equation}
with $(V^b,V^a)=(V_0,V_0)$ for $\alpha=0$ and $(V^b,V^a)=(\vlrg,\vsml)$ for $\alpha\in\{\uparrow,\downarrow\}$. In particular, the integral above is invariant under the exchange $V^b\leftrightarrow V^a$, so that
\begin{equation}
   \langle t_{hit,\uparrow}\rangle,  \langle t_{hit,\downarrow}\rangle \xrightarrow[Q\to\infty]{} \langle t_{hit}\rangle
\end{equation}
We remark that the $Q\to\infty$ limit of $\langle t_{hit}\rangle$ for $\mu>0$ is independent of $Q$, so that the asymptotic scaling of the average execution time is dominated by the exponential divergence of the hitting number. Also notice that, even though the time required to hit the $V^a=Q$ boundary is also diverging at a speed proportional to $Q$, its contribution is subleading: The barrier $V^a=Q$ needs to be hit just once, as opposed to the $V^a,V^b=0$ boundaries which are hit a large number of times. \\
The phenomenology of the case $\mu=0$ is extremely different: while the hitting number scales like $Q^2$, the average $\langle t_{hit,\alpha}\rangle$ is proportional to $\log Q$, implying that the average execution time scales as $Q^2\log Q$.
Indeed, in both cases the increase of the expectation time with the volume justifies the large volatility risk which a trader faces in the large $Q$ regime.\\
In limit of small executed volumes, the average execution time is dominated by
\begin{equation}
  \langle T \rangle \xrightarrow[Q\to V_0]{} \langle t_{hit,0} \rangle \; ,
\end{equation}
where $t_{hit,0}$ can be expanded around $Q=V_0$ by using the results of App.~\ref{app:tot_hit_time}, allowing to express the average execution time as an increasing function of the difference $Q-V_0$.

\section{Discussion and extensions}
\label{sec:discussion_and_extensions}
The model that we have presented deliberately simplifies the rich structure of a real order book, and does not take into account several of its microscopic features. We believe, motivated by the result of numerical simulations and the inspection of empirical data, that the features neglected in the current version of the model can be progressively reintroduced without modifying its main qualitative predictions. In particular, the essential feature on which our model relies is the fact that any asymmetry in liquidity (i.e., imbalances between bid and ask volumes) induces a corresponding asymmetry in the future direction of price, a well-known stylized fact of market microstructure summarized by the statement that ``(efficient) price is where the volume is not''~\cite{Cont:2013ab,delattre2013estimating}. This is why we believe our findings to be robust.

It would nevertheless be interesting to encode some more realistic features in the model, so to obtain predictions quantitatively more accurate. In particular:
\begin{enumerate}
    \item A stochastic volume after the depletion of the queues (as opposed to the deterministic values $\vsml$ and $\vlrg$) can be introduced with little expense. The only change required in our equations is the substitution
    \begin{equation}
        p_{\alpha}(t,V^b,V^a) \to \int dV^b dV^a p_{\alpha}(t,V^b,V^a) \pi_\pm(V^b,V^a) \, ,
    \end{equation}
    where $\pi_\pm(V_b,V_a)$ refers to the distribution of the volumes after the queues reset. Hence, the analytical results presented above allow to solve the model even in this more general setting.
    \item The dynamics of the best queues in an actual order book evolves discontinuously through jumps in volume, consistently with the description adopted in~\cite{Cont:2010aa,Cont:2013aa}. Using Poissonian jumps of the queues allows to capture this feature, at the expense of a more involved analytical procedure required to obtain the first-passage probabilities on a semi-infinite domain. While this modification can be substantial for many contracts, when the tick size is sufficiently large the neglection of this effect becomes progressively less relevant: for liquid stocks in equity markets, the typical duration between events at the best queue is of the order of the tens of milliseconds, while when executing an order, the focus is typically on times larger then some seconds~\cite{cont2012order}.
    \item The assumption of independence for the Fokker-Planck equations for the two queues is a rather drastic simplification (dropped, for example, in~\cite{cont2012order}). Moreover, the empirical results of Ref.~\cite{Gareche:2013aa} indicate the actual Fokker-Planck equation describing the diffusive limit of the queues dynamics has an even more complex structure than the one considered in~\cite{cont2012order}. Keeping this effect into account is likely to force one to calculate the absorption probabilities numerically, due to the involved structure of the resulting Fokker-Planck equation.
    \item The effect of a bid-ask asymmetry in the parameters of the model, as considered in~\cite{cont2012order}, can also be addressed, so to model for example the presence of local price trends.
\end{enumerate}
All these extensions modify the absorption probabilities $p_{\alpha,\alpha^\prime}$ while leaving invariant the expressions for the generating functions $\Phi_x(\omega,s),\Phi(\omega,s),\Psi_x(\omega,s),\Psi_n(\omega,s)$, which can then be employed in order to study any Markovian model in which the queues are reset after exhaustion. In particular, when closed-form expressions for the absorption probabilities are not available, it is possible to compute them via Monte Carlo. This allows to relate the empirical results concerning the time of exhaustion of the queues with the expected price change following a one-shot execution.

\begin{figure}[h]
    \begin{center}
        \includegraphics{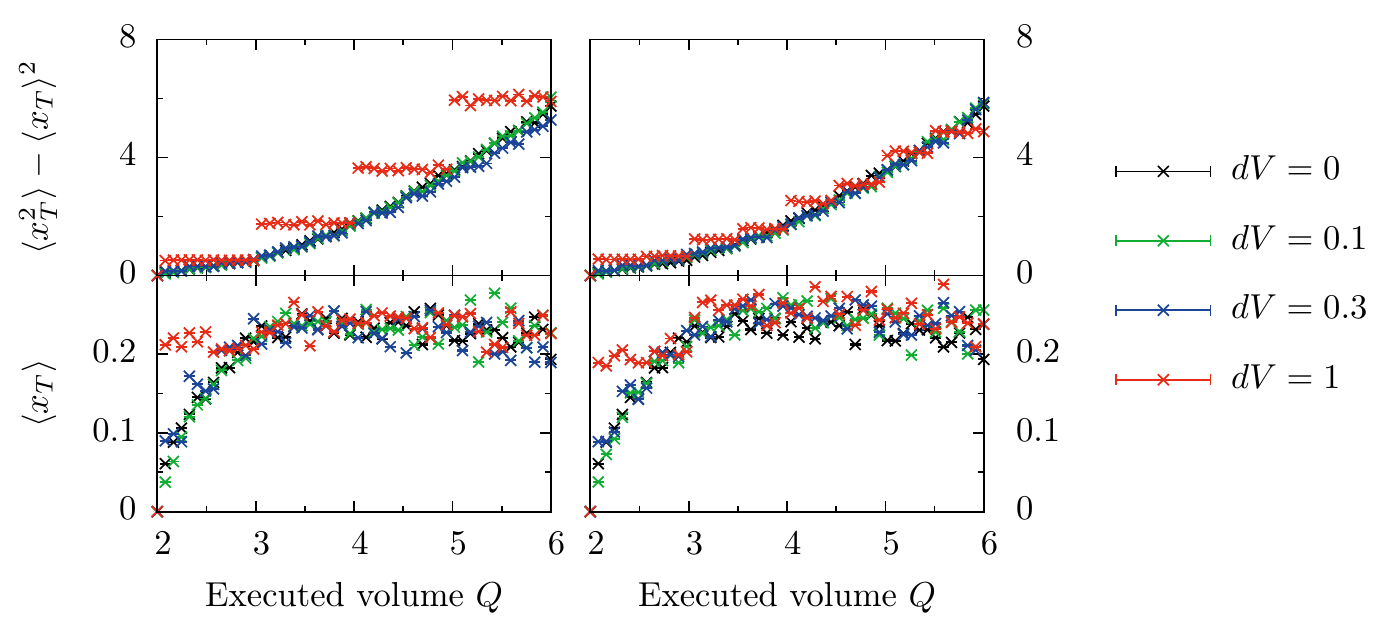}
    \end{center}
    \caption{Average price changes $\langle x_T \rangle$ and price fluctuations $\langle x^2_T \rangle-\langle x_T \rangle^2$ following the execution of a trade of variable volume $Q$. The left plot shows the effect of granularity in time and volume, by considering the queue volumes to be subject to jumps of discrete volume at random Poissonian time. We fixed the parameters of the model in order to match $\mu=0$ and $D=1$ in the continuous-volume limit of the model. As in Fig.~\ref{fig:exec}, we have taken $\vsml=1$ and $\vlrg=3$. We have used here a parameter $d V$ interpolating among the continuous case $d V = 0$ and the rougher case $d V =1$. The right plot superposes to this effect the one of a stochastic volume for the queues after reset ($\vlrg$ uniformly distributed in $\{2,2.5,3,3.5,4\}$ and $\vsml$ uniformly distributed in $\{ 0.5,1,1.5\}$). The plots result from 5000 realizations of the execution process, and suggest that the predictions of the model are robust with respect to the effects described in Sec~\ref{sec:discussion_and_extensions}: starting from $dV=0.3$, the curves are almost superposed to the ones obtained in the continuous volume case.}
    \label{fig:gen_exec}
\end{figure}

In Fig.~\ref{fig:gen_exec} we investigate how our results vary by introducing granularity both in time and volume (we consider Poissonian jumps of discrete size), and by allowing the volumes of the queues after reset to fluctuate.

The key technical assumption of our approach is the one of Markovianity, which we need in order to be able to solve the model. While this assumption is justified for large tick contracts~\cite{Cont:2013aa}, in order to model products of progressively smaller tick it would be necessary to explore the spacial structure of the book by enlarging the number of price levels around the best price included in the description.

\section{Conclusions}
\label{sec:concl}
This work shows that even in a risk-neutral setting and in absence of a price trend, the market impact of an order cannot be completely avoided by delaying the transaction: even if a one-shot strategy is adopted, thus completely removing the traditional notion of price impact from the cost of the execution strategy, such a cost reappears due to an asymmetric conditioning effect induced by the execution strategy. Moreover, the volatility risk associated with such a passive strategy can be (exponentially) large if the volume to execute is too ingent. This indicates that the access to liquidity has a price which is to some extent unavoidable. Within our analysis we have characterized such cost, its fluctuations, and provided the probability of completely executing a one-shot order in a stylized market model, relating these quantities to microstructural parameters.
Even though we have expressed our ideas in a rather ad-hoc setting, we want to underline that even in a more general scenario these qualitative predictions are expected to be observed: any type of correlation among direction of price changes and liquidity may cause asymmetric effects which can induce an apparent impact effect for passive traders. In particular, as long as asymmetry in the volumes of the best queues are predictive for the direction of future price changes, we expect this effect to be detected.
Notice also that the presence of opportunistic, one-shot traders would induce a liquidity-induced mean reversion effect in the price due to their tendency to exhaust large volume queues, which typically appear on the side of the market which is opposite to the one where the price is currently trending.

As a future research perspective, we look forward for the possibility of validating the qualitative predictions of this model through an inspection of proprietary data describing the execution of one-shot orders.

\section*{Acknowledgements}
The author acknowledges interesting discussions with J.-P.~Bouchaud, N.~Cosson and B.~T\'oth during the preparation of this manuscript.
This research benefited from the support of the ``Chair Markets in Transition'', under the aegis of ``Louis Bachelier Finance and Sustainable Growth'' laboratory, a joint initiative of \'Ecole Polytechnique, Universit\'e d'\'Evry Val d'Essonne and F\'ed\'eration Bancaire Fran\c{c}aise.

\appendix

\section{Diffusion in a semi-infinite strip}
\label{app:prob_abs}
In this appendix we will be interested in computing the probabilities $\pex(t,V^b,V^a)$, $\pup(t,V^b,V^a)$, $\pdn(t,V^b,V^a)$ introduced in Sec.~\ref{sec:model}, denoting respectively the probability of executing a trade of volume $Q$, emptying the ask queue and emptying the bid queue in a time $t$ starting from bid and ask volumes respectively equal to $V^b$ and $V^a$.
We will focus exclusively on the case in which a trader is present, reminding that free regime discussed in Sec.~\ref{sec:free} can be recovered by taking the limit  $Q\to\infty$, in which the execution probability $\pex(t,V^b,V^a)$ becomes zero.
Finally, we remind that this problem corresponds to the one of finding the first-passage probabilities of a diffusing particle through any of the three boundaries of the semi-infinite strip $(0,\infty)\times(0,Q)$.

Such first-passage probabilities $\hat p_\alpha$ satisfy the set of the independent equations
\begin{equation}
  \label{eq:helmholtz}
  \frac{1}{2} \bm{\nabla}^2 \hat p_\alpha (\omega,\vecv) -(\bm{\nabla}\cdot\bm{\mu})\, \hat p_\alpha(\omega,\vecv) \cdot = \omega \hat p_\alpha(\omega,\vecv)
\end{equation}
where $\bm{\nabla}=(\partial_x,\partial_y)$, $\bm{\mu}=(\mu,\mu)$, $\vecv=(V^b,V^a)$ and $\alpha \in \{ ex,\uparrow,\downarrow\}$. These equations differ for the boundary conditions as specified in Sec.~\ref{sec:model}. Specifically, $p_\alpha(\omega,\vecv)$ is equal to zero on all the boundaries of the region $(0,\infty)\times(0,Q)$ except for the one corresponding to the event labeled by $\alpha$ (i.e., $\hex = 1$ for $V^a=Q$, $\hup = 1$ for $V^a=0$, $\hdn = 1$ for $V^b=0$). The choice $\omega=0$ is associated to the probability of hitting a specific boundary in any point in time, in whose case the problem reduces to a simpler problem, for which we will be able to provide explicit solutions. \\
In order to solve this problem with elementary methods, we define the functions $\phi_\alpha(\omega,\vecv)= \exp(-\mu(V^a+V^b)) \hat p_\alpha(\omega,\vecv)$, which satisfy the simpler set of Helmholtz equations
\begin{equation}
  \label{eq:red_helmholtz}
   \bm{\nabla}^2 \phi_\alpha (\omega,\vecv)  = 2 (\omega + \mu^2) \phi_\alpha
\end{equation}
subject to the modified boundary conditions
\begin{eqnarray}
  \label{eq:red_bnd}
  \phi_{ex}(\omega,V^b,Q) &=& e^{-\mu (V^a+Q)} \\
  \phi_\uparrow(\omega,V^b,0) &=& e^{-\mu V^b} \\
  \phi_\downarrow(\omega,0,V^a) &=& e^{-\mu V^a}  \;.
\end{eqnarray}
The problem of determining $\phi_\alpha$ is more conveniently  handled by  treating the semi-infinite strip $(0,\infty)\times(0,Q)$ as the $P\to\infty$ limit of the rectangle $(0,P)\times(0,Q)$, in whose geometry the general solution of the problem~(\ref{eq:helmholtz}) can be written as follows:
\begin{eqnarray}
  \label{eq:9}
 \phi_{ex}(\omega,\vecv) &=&  -\int_0^P d\zeta \left[ \frac{\partial}{\partial \eta} G(\vecv,\zeta,\eta) \right]_{\eta=Q}
\phi_{ex}(\omega,\zeta,Q) \\
  \phi_\uparrow(\omega,\vecv) &=&\phantom{-} \int_0^P d\zeta \left[ \frac{\partial}{\partial \eta} G(\vecv,\zeta,\eta) \right]_{\eta=0}
\phi_\uparrow(\omega,\zeta,0)  \\
  \phi_\downarrow(\omega,\vecv) &=&\phantom{-} \int_0^Q d\eta \left[ \frac{\partial}{\partial \zeta} G(\vecv,\zeta,\eta) \right]_{\zeta=0}
\phi_\downarrow(\omega,0,\eta)
\end{eqnarray}
where the Green function $G(\vecv,\zeta,\eta)$ admits the two forms
\begin{eqnarray}
  \label{eq:19}
  G(\vecv,\zeta,\eta) &=& \frac{2}{P} \sum_{n=1}^\infty \frac{\sin (p_n V^b)\sin(p_n \zeta)}{\beta_n \sinh (\beta_n Q)} H_n(V^a,\eta) \\
&=& \frac{2}{Q} \sum_{n=1}^\infty \frac{\sin (q_n V^b)\sin(q_n \zeta)}{\mu_n \sinh (\mu_n a)} Q_n(V^b,\zeta) \;,
\end{eqnarray}
and where we have defined
\begin{eqnarray}
  \label{eq:20}
  p_n = \frac{\pi n}{P} \quad \; &,&  \quad 
  \beta_n = \sqrt{p_n^2 + 2 \omega +2\mu^2} \; , \\
  q_n  =  \frac{\pi n}{Q} \quad \; &,& \quad
  \mu_n =\sqrt{q_n^2 + 2 \omega +2\mu^2} \; .
\end{eqnarray}
Finally,
\begin{eqnarray}
  \label{eq:46}
  H_n(V^a,\eta) &=& \left\{
  \begin{array}{ccc}
    \sinh (\beta_n \eta) \sinh (\beta_n(Q-V^a)) &\textrm{ if }& V^a>\eta \\
    \sinh (\beta_n V^a) \sinh (\beta_n(Q-\eta)) &\textrm{ if }& \eta > V^a
  \end{array}
  \right.
\\
  Q_n(V^b,\zeta) &=&\left\{
  \begin{array}{ccc}
    \sinh (\mu_n \zeta) \sinh (\mu_n(P-V^b)) &\textrm{ if }& V^b>\zeta  \\
    \sinh (\mu_n V^b) \sinh (\mu_n(P-\zeta)) &\textrm{ if }& \zeta>V^b
  \end{array}
  \right.
\end{eqnarray}
The explicit form of the solution for the $\phi_\alpha$ can be written after performing the above integrals and taking the $P\to\infty$ limit.  A further transformation back to the original functions $\hat p_\alpha$ finally allows to express the solution as the series
\begin{eqnarray}
  \hex (\omega,\vecv) &=&  \frac{2}{\pi} \sum_{n} \frac{\sin(q_n V^a)}{n} \frac{(-1)^{n+1}e^{-\mu(Q- V^a)}}{1+(2\omega +\mu^2)Q^2/\pi^2 n^2} (1-e^{-(\mu_n-\mu)\;V^b}) \\
  \hup (\omega,\vecv) &=& \frac{2}{\pi} \sum_{n} \frac{\sin(q_n V^a)}{n} \frac{e^{\mu V^a}}{1+(2\omega +\mu^2)Q^2/\pi^2 n^2} (1-e^{-(\mu_n-\mu)\;V^b})  \\
  \hdn (\omega,\vecv) &=& \frac{2}{\pi} \sum_{n} \frac{\sin(q_n V^a)}{n} \frac{e^{\mu V^a}}{1 +\mu^2Q^2/\pi^2 n^2} (1-(-1)^n e^{-\mu Q}) e^{-(\mu_n-\mu)\;V^b} \; ,
\end{eqnarray}
or equivalently as the integral
\begin{eqnarray}
  \label{eq:int_rep_hex}
  \hex (\omega,\vecv) &=& \frac{2}{\pi} \int_0^\infty  dp \frac{p}{p^2+\mu^2} \frac{\sin(p V^b) \sinh(\beta(p) V^a)}{\sinh(\beta(p) Q)}e^{\mu V^b -\mu(Q-V^a)} \\
  \label{eq:int_rep_hup}  
  \hup (\omega,\vecv) &=& \frac{2}{\pi} \int_0^\infty dp  \frac{p}{p^2+\mu^2} \frac{\sin(p V^b) \sinh(\beta(p) (Q-V^a))}{\sinh(\beta(p) Q)} e^{\mu V^b + \mu V^a} \\
  \label{eq:int_rep_hdn}
\hdn (\omega,\vecv) &=& \frac{2}{\pi} \int_0^\infty  dp \frac{p}{p^2+\mu^2+2\omega} \left( \frac{\sin(pV^b)}{\sinh(\beta(p)Q)}\right) e^{\mu V^b} \\
&\times& \bigg( \sinh(\beta(p) Q) - e^{ - \mu(Q-V^a)} \sinh(\beta(p)V^a ) - e^{ \mu V^a}\sinh(\beta(p)(Q-V^a) ) \bigg) \nonumber \; .  
\end{eqnarray}

\subsection{Hitting probabilities in the free, driftless case}
In the special case $\mu=\omega=0$ it is possible to sum analytically the above series so to express the result in term of elementary functions. It is sufficient to remind that
\begin{equation}
  \label{eq:log_series}
  \sum_{n=1}^{\infty} (-1)^{n+1} \frac{e^{zn}}{n} = \log (1+e^{zn}) 
\end{equation}
and to expand the trigonometric functions in term of exponentials in order to reduce the above series to sums of logarithms. Then, by exploiting the identities
\begin{equation}
  \frac{i}{2} \left( \log (1 \pm e^{i\pi V^a /Q}e^{-\pi V^b/Q})- \log (1 \pm e^{-i\pi V^a /Q}e^{-\pi V^b/Q}) \right) = \arctan\left( \frac{\sin(\pi V^a/Q)}{e^{\pi V^b /Q} \pm \cos(\pi V^a /Q)} \right)
\end{equation}
and
\begin{equation}
\frac 2 \pi  \sum_{n=1}^{\infty} \frac{\sin(\pi V^a/Q)}{n} = 1- \frac{V^a}{Q} 
\end{equation}
one can obtain the following expression for the hitting probabilities:
\begin{eqnarray}
  \label{eq:hex_dless_om_0}
  \bex(\vecv) &=& \frac{V^a}{Q} - \frac{2}{\pi} \arctan\left( \frac{\sin(\pi V^a/Q)}{e^{\pi V^b / Q} + \cos(\pi V^a / Q)}\right) \\
  \label{eq:hup_dless_om_0}
  \bup(\vecv) &=& 1 - \frac{V^a}{Q}  - \frac{2}{\pi} \arctan\left( \frac{\sin(\pi V^a/Q)}{e^{\pi V^b / Q} - \cos(\pi V^a / Q)}\right)\\
  \label{eq:hdn_dless_om_0}
  \bdn(\vecv) &=& \frac{2}{\pi} \arctan\left( \frac{\sin(\pi V^a/Q)}{e^{\pi V^b / Q} - \cos(\pi V^a / Q)}\right) \nonumber \\
&+&
\frac{2}{\pi} \arctan\left( \frac{\sin(\pi V^a/Q)}{e^{\pi V^b / Q} + \cos(\pi V^a / Q)}\right) \; .
\end{eqnarray}
Interestingly enough, the dependence on $\omega$ of the absorption probabilities close to the point $\omega=0$ is regular enough to lead to finite mean hitting times for any of the boundaries, as opposed to the $\mu=0$ case of the free diffusion problem, in which all these quantities were divergent.

\subsection{Small $\omega$ expansion in the free case}
The asymptotic analysis of Eqs.~(\ref{eq:int_rep_hex}),~(\ref{eq:int_rep_hup}) and~(\ref{eq:int_rep_hdn}) is particularly interesting, and has been used in order to determine the large time behavior of the free diffusion problem (corresponding to the $Q=\infty$, $\omega\to 0$ regime) in Sec.~\ref{sec:model}. Such behavior depends crucially upon $\mu$: for $\mu>0$ the hitting probabilities are analytic around $\omega=0$, and thus can be expanded as
\begin{equation}
  \label{eq:free_exp_drift}
  \hat p_{\alpha}(\omega,\vecv) = \bar p_{\alpha}(\vecv) \left( 1 - \omega \langle t_\alpha (\vecv)\rangle + \frac{1}{2} \omega^2 \langle t_\alpha^2(\vecv)\rangle  + O(\omega^3)
  \right)
\end{equation}
where the symbols $\langle t_{\alpha}(\vecv)\rangle$ and $\langle t^2_{\alpha}(\vecv)\rangle$ refer to the average hitting times conditional to the initial condition $\vecv=(V^b,V^a)$ and a first absorption though a boundary of type $\alpha$. An integral representation these terms can be obtained straightforwardly by differentiation with respect to $\omega$ under the integral sign.\\
For $\mu=0$ the hitting probabilities are non-analytic around $\omega=0$, and admit in particular the expansion
\begin{equation}
  \label{eq:free_exp_dless}
  \hat p_{\alpha}(\omega,\vecv) = \sum_n a_{\alpha}^{(n)}(\vecv) \omega^n + \log (\omega) \sum_n b_{\alpha}^{(n)}(\vecv) \omega^n \; , 
\end{equation}
with $a_\alpha^{(0)} = \bar p_{\alpha}(0)$ and $b_{\alpha}^{(0)} = 0$. The subleading terms in the expansion are
\begin{eqnarray}
 \label{eq:free_exp_dless_coeff_a}
 a^{(1)}(\vecv) &=& \frac{V^bV^a}{\pi} \left( (2\gamma-3) + 2\frac{V^a}{V^b} \arctan \left( \frac{V^b}{V^a} \right) + \log\left( \frac{(V^b)^2 + (V^a)^2}{2} \right) \right) 
\\
 \label{eq:free_exp_dless_coeff_b}
 b^{(1)}(\vecv) &=& \frac{V^bV^a}{\pi} \; ,
\end{eqnarray}
where $\gamma$ denotes the Euler-Mascheroni constant. These terms can be obtained by using the series representation for the absorption probabilities
\begin{equation}
  \hup (\omega,\vecv) =\frac{2}{\pi} \sum_{n=0}^\infty \frac{(-1)^n}{(2n+1)!} \left( \frac{V^b}{V^a} \right)^{2n+1} \int_0^\infty z^{2n} e^{- \sqrt{z^2+2\omega (V^a)^2}} \; .
\end{equation}
The correction terms~(\ref{eq:free_exp_dless_coeff_a}) and~(\ref{eq:free_exp_dless_coeff_b}) are finally recovered after exploiting the identity $\int_0^\infty z^{2n} e^{- \sqrt{z^2+\alpha^2}} = (2n-1)!! \alpha^{n+1} K_{n+1}(\alpha)$, with $K_{n+1}(\alpha)$ denoting the modified Bessel function of the second type of order $n+1$ calculated in $\alpha$. Notice in particular that the divergence of the Bessel function close to $\alpha=0$ is canceled by the factor $\alpha^{n+1}$.

\subsection{Hitting probabilities in the large $Q$ regime}
The regime of large $Q$ and $\omega=0$ has been analyzed in order to find the asymptotic scaling of the price change $x_t$ and of the hitting number $n_t$ in the execution problem illustrated in Sec.~\ref{sec:one_shot}. In the case $\mu> 0$, Eqs.~(\ref{eq:int_rep_hex}),~(\ref{eq:int_rep_hup}) and~(\ref{eq:int_rep_hdn}) can be expanded as
\begin{eqnarray}
  \label{eq:hex_lrg_Q_drift}
  \bex (\vecv,Q) &\xrightarrow[Q\to\infty]{}& \bex (\vecv,\infty) +\frac{x}{\sqrt{\pi } \mu^2} \left(\frac{2\sqrt{2} \mu }{ Q}\right)^{3/2} \sinh \left(\sqrt{2} \mu  y\right) e^{\mu  (x+y-(1+\sqrt{2})Q)} \\
  \label{eq:hup_lrg_Q_drift}
  \bup (\vecv,Q) &\xrightarrow[Q\to\infty]{}& \bup (\vecv,\infty) -\frac{x}{\sqrt{\pi } \mu^2} \left(\frac{\sqrt{2} \mu }{ Q}\right)^{3/2} \sinh \left(\sqrt{2} \mu  y\right) e^{\mu  (x+y-2\sqrt{2}Q)}\\
  \label{eq:hdn_lrg_Q_drift}
  \bdn (\vecv,Q) &\xrightarrow[Q\to\infty]{}& \bdn (\vecv,\infty) -\frac{x}{\sqrt{\pi } \mu^2} \left(\frac{2\sqrt{2} \mu }{ Q}\right)^{3/2} \sinh \left(\sqrt{2} \mu  y\right) e^{\mu  (x+y-(1+\sqrt{2})Q)} ,
\end{eqnarray}
while for $\mu=0$ one can simply differentiate Eqs.~(\ref{eq:hex_dless_om_0}),~(\ref{eq:hup_dless_om_0}) and~(\ref{eq:hdn_dless_om_0}) and obtain
\begin{eqnarray}
  \label{eq:hex_lrg_Q_dless}
  \bex (\vecv,Q) &\xrightarrow[Q\to\infty]{}& \bex (\vecv,\infty) + \frac{\pi V^aV^b}{2Q^2}\\
  \label{eq:hup_lrg_Q_dless}
  \bup (\vecv,Q) &\xrightarrow[Q\to\infty]{}& \bup (\vecv,\infty) -\frac{\pi V^aV^b}{6Q^2}\\
  \label{eq:hdn_lrg_Q_dless}
  \bdn (\vecv,Q) &\xrightarrow[Q\to\infty]{}& \bdn (\vecv,\infty) -\frac{\pi V^aV^b}{3Q^2} \; .
\end{eqnarray}
Above expressions can be inserted in the formulae for $\langle x_T\rangle$ and $\langle n_T\rangle$ calculated in App.~\ref{app:markov_chain} so to estimate their large $Q$ scaling.

\subsection{Unconditional hitting time}
\label{app:tot_hit_time}
The unconditional hitting time $t_{hit}(\vecv)$, measuring the average time required to hit any of the boundaries, is defined in App.\ref{app:markov_chain}, where its averages are expressed as
\begin{equation}
  \langle t_{hit}(\vecv)\rangle = - \frac{d}{d\omega} (\hex + \hup + \hdn) \bigg|_{\omega=0}\;,
\end{equation}
We want to show that it can be conveniently computed by first summing Eqs.~(\ref{eq:int_rep_hex}),~(\ref{eq:int_rep_hup}) and~(\ref{eq:int_rep_hdn}), and successively performing the derivative with respect to $\omega$, in order to exploit the cancellations among the summands. We obtain in particular
\begin{eqnarray}
  \label{eq:tot_hit_time}
  \langle t_{hit}(\vecv)\rangle &=& \frac{4}{\pi} \int_0^\infty dp \frac{p}{(p^2+\mu^2)^2} \sin(p V^b) \\
&\times& \left( e^{\mu V^b}- \frac{\sinh(\beta(p)V^a) }{\sinh(\beta(p)Q )}e^{\mu(V^b+V^a-Q)}  -\frac{\sinh(\beta(p)(Q-V^a))  }{\sinh(\beta(p)Q )}e^{\mu(V^b+V^a)} \right) \; , \nonumber
\end{eqnarray}
which in the driftless case $\mu=0$ reduces to
\begin{equation}
  \label{eq:tot_hit_time_dless}
  \langle t_{hit}(\vecv)\rangle = \frac{4Q^2}{\pi}  \int_0^\infty dp \frac{\sin(p V^b/Q) }{p^3} \left[ 1-\cosh(pV^a/Q) + \tanh(p/2) \sinh(pV^a/Q) \right] \; .
\end{equation}

\section{Solution of a Markovian market model}
\label{app:markov_chain}
\subsection{Generating functions for the free case}
\begin{figure}[h]
   \centering
   \includegraphics[width=0.9\textwidth]{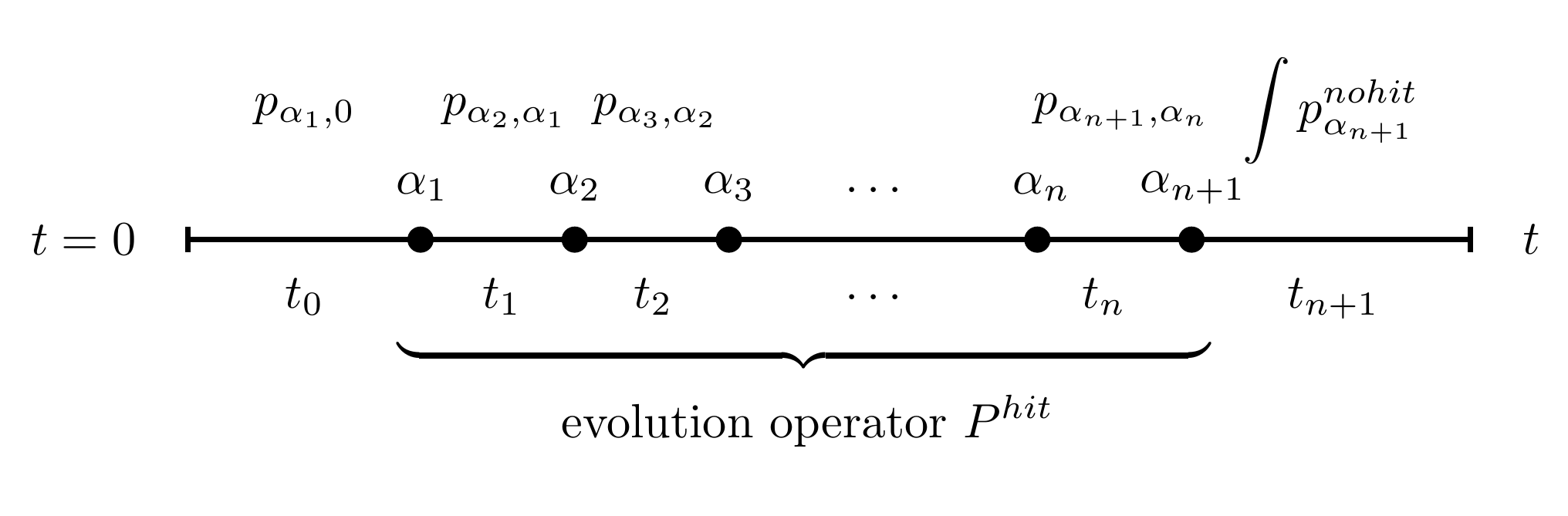} 
   \caption{Schematic representation of a generic term of the evolution equation~(\ref{eq:markovian_evol}) for the problem of the free evolution of the price.}
   \label{fig:markovian_scheme}
\end{figure}

In Sec.~\ref{sec:model} we have considered the problem of calculating the quantities $P_x(t,x)$ and $P_n(t,n)$ describing the probability for the price and the hitting number to take values respectively of $x$ and $n$ at time $t$. The former one can be written as
\begin{eqnarray}
\label{eq:markovian_evol}
 P_x(t,x)  &=& \sum_{n=0}^\infty \sum_{\alpha_1}\dots \sum_{\alpha_{n+1}} \int_0^\infty \left(\, \prod_{i=0}^{n+1} dt_i\, \right)  \delta\left(t-\sum_{i=0}^{n+1} t_i\right) \delta\left(x - \sum_{i=1}^{n+1} X(\alpha_i)\right) \nonumber \\
&\times& \left( \int_0^{t_{n+1}} dt^\prime\, (1-p_{\uparrow \alpha_{n+1}}(t^\prime) -p_{\downarrow \alpha_{n+1}}(t^\prime)) \right) \left(\, \prod_{i=1}^n p_{\alpha_{i+1} \alpha_i}(t_i) \, \right) p_{\alpha_1 0} \nonumber \\
&+& \int_0^t dt_0 \, (1-p_{\uparrow 0}(t) -p_{\downarrow 0}(t)) \, \delta(x)  \; ,
\label{eq:free_prob_price}
\end{eqnarray}
where $X(\uparrow)=-X(\downarrow)=1$. Each of the terms in above equation  can be represented schematically as in Fig.~\ref{fig:markovian_scheme}, in which we show that the $n$-th term of the sum comprises $n+1$ absorptions from the starting time to the last one $t$, while the succession of indexes $\{ \alpha_i \}_{i=1}^{n+1} \in \{ \uparrow,\downarrow\}^{n+1}$ labels the type of boundary hit during absorption number $i$. The probability $P_n(n,t)$ can be obtained from the previous expression after the substitution $X(\alpha_i) \to 1$.
The generating function associated to the probability distribution~(\ref{eq:free_prob_price}) is readily found by integrating in the $t$ and $x$ coordinates, and can be expressed compactly in matrix form as in Eq.~(\ref{eq:free_gf_price}). The generating function for the hitting number is found analogously, leading to Eq.~(\ref{eq:free_gf_hit}). We can rewrite Eqs.~(\ref{eq:free_gf_price}) and~(\ref{eq:free_gf_hit}) more succinctly as
\begin{eqnarray}
\label{eq:free_gf_price_app}
\Phi_x(\omega,s) &=& \frac 1 \omega \left[ \hat p_0^{nohit} + (\hat{\bm{p}}^{nohit})^T \sum_{n=0}^\infty (\hat P^{hit} K_x)^n  K_x\hat{\bm{p}}^{hit}_0 \right] \\
\label{eq:free_gf_hit_app}
\Phi_n(\omega,s) &=& \frac 1 \omega \left[ \hat p_0^{nohit} + (\hat{\bm{p}}^{nohit})^T \sum_{n=0}^\infty (\hat P^{hit} K_n)^n  K_n\hat{\bm{p}}^{hit}_0 \right] \; ,
\end{eqnarray}
where we have defined the matrices
\begin{equation}
\hat P^{hit}=
\left(
  \begin{array}{cc}
    \huu & \hud  \\
    \hdu & \hdd
  \end{array}
\right)
\quad 
K_x=
\left(
  \begin{array}{cc}
    e^{-s} & 0  \\
    0 & e^s
  \end{array}
\right)
\quad
K_n=
\left(
  \begin{array}{cc}
    e^{-s} & 0 \\
    0 & e^{-s}
  \end{array}
\right) \; ,
\end{equation}
together with the vectors
\begin{equation}
  \hat{\bm{p}}^{\;hit}_0 =
\left(
  \begin{array}{c}
    \huz \\
    \hdz
  \end{array}
\right)
\quad
 \hat{\bm{p}}^{\;nohit} =
\left(
  \begin{array}{c}
    1-\huu - \hdu  \\
    1-\hud - \hdd
  \end{array}
\right)
\end{equation}
and the scalar $\hat p^{nohit}_0 = 1-\huz -\hdz$.
Notice the factor $\omega^{-1}$ appearing in Eqs.~(\ref{eq:free_gf_price_app}) and~(\ref{eq:free_gf_hit_app}), which arises from the integration by parts of the last time $t_{n+1}$. The infinite sum appearing in the expression for the generating functions can be performed explicitly by passing to principal component. One obtains
\begin{eqnarray}
\Phi_x(\omega,s) &=& \frac 1 \omega \left[ \hat{p}_0^{nohit} + (\hat{\bm{p}}^{nohit})^T U_x (\mathbb{I} - \Lambda_x)^{-1}\, U^{-1}_x\,\hat{\bm{p}}^{hit}_0 \right] \\
\Phi_n(\omega,s) &=& \frac 1 \omega \left[ \hat{p}_0^{nohit} + (\hat{\bm{p}}^{nohit})^T U_n (\mathbb{I} - \Lambda_n  )^{-1}\,U^{-1}_n\,  \hat{\bm{p}}^{hit}_0 \right] \; ,
\end{eqnarray}
where $\mathbb{I}$ denotes the identity matrix, and where we have introduced the eigenvalue decompositions $\hat P^{hit} K_x = U_x \Lambda_x U^{-1}_x$ and $\hat P^{hit} K_n = U_n \Lambda_n U^{-1}_n$. The matrix product appearing in the first of above equations results
\begin{equation}
U_x (\mathbb{I} - \Lambda_x)^{-1}\, U^{-1}_x =
\frac 1 {1- \hud\hdu + \huu\hdd - e^{-s}\huu - e^s\hdd}
\left(
\begin{array}{cc}
  1-e^s\hdd  & e^{-s}\hud \\
  e^s \hdu & 1- e^{-s}\huu
\end{array}
\right)
\; ,
\end{equation}
while the one relative to the second equation can be obtained through the substitution $\hat p_{\downarrow\alpha}\to e^{-2s} \hat p_{\downarrow\alpha}$.

The differentiation of the generating functions with respect to $s$ leads finally to the Laplace transforms of mean price change $\langle\hat x_\omega \rangle$ and hitting number $\langle\hat n_\omega \rangle$, together with their squares $\langle\hat x_\omega^2 \rangle$ and $\langle\hat n^2_\omega \rangle$:
\begin{eqnarray}
  \label{eq:free_av_price}
  \langle \hat x_\omega \rangle &=& \frac 1 \omega \left( \frac
    {\huz (1- \hdd - \hdu) - \hdz (1-\huu-\hud)}
    {1- \hud\hdu + \huu\hdd - \huu - \hdd} \right)
\\
  \label{eq:free_sq_price}
  \langle \hat x^2_\omega \rangle &=& \frac \huz \omega \left[ \frac{
 (1-2\hdd +\hdd^2 -\hdu + 3\hdd\hdu -\hdu \hud -\hdd \hdu \hud )
}{
(1-\hdd-\huu-\hud\hdu+\huu \hdd )^2
} \right. \\ \nonumber
&+& \left.
\frac{ 
\hdu^2 \hud + \huu - 2 \hdd \huu + \hdd^2 \huu - \hdu \huu -\hdd \hdu \huu
}{
(1-\hdd-\huu-\hud\hdu+\huu \hdd )^2 
} \right] \\
 &+& \textrm{\large (} \uparrow \textrm{ exch. } \downarrow \textrm{\large )} \nonumber
\\ \nonumber \\
  \label{eq:free_av_hit}
  \langle \hat n_\omega \rangle &=& \frac 1 \omega \left( \frac
    {\huz (1- \huu + \hdu) + \hdz (1-\huu+\hud)}
    {1- \hud\hdu + \huu\hdd - \huu - \hdd} \right)
\\
  \label{eq:free_sq_hit}
  \langle \hat n^2_\omega \rangle &=& \frac \hup \omega \left[ \frac{
 (1-2\hdd +\hdd^2 +3\hdu - \hdd\hdu +3 \hdu \hud -\hdd \hdu \hud )
}{
(1-\hdd-\huu-\hud\hdu+\huu \hdd )^2
} \right. \\ \nonumber
&+& \left.
\frac{ 
\hdu^2 \hud + \huu - 2 \hdd \huu + \hdd^2 \huu - \hdu \huu -\hdd \hdu \huu
}{
(1-\hdd-\huu-\hud\hdu+\huu \hdd )^2 
} \right] \\
 &+& \textrm{(} \uparrow \textrm{ exch. } \downarrow \textrm{)} \nonumber
 \; .
\end{eqnarray}

\subsection{Generating functions for the one-shot execution problem}
The problem which we are required to solve during the execution of a one-shot order requires finding the generating functions for the probabilities $P^{ex}_x(T,x)$ and $P^{ex}_n(T,x)$ that the price and the hitting number assume respectively the values of $x$ and $n$ at the stopping time $T$. Those probabilities follow a law extremely similar to Eq.~(\ref{eq:free_prob_price}), we have in fact
\begin{eqnarray}
 P^{ex}_x(t,x)  &=& \sum_{n=0}^\infty \sum_{\alpha_1}\dots \sum_{\alpha_{n+1}} \int_0^\infty \left(\, \prod_{i=0}^{n+1} dt_i\, \right)  \delta\left(t-\sum_{i=0}^{n+1} t_i\right) \delta\left(x - \sum_{i=1}^{n+1} X(\alpha_i)\right) \nonumber \\
&\times& p_{ex,\alpha_{n+1}}(t_{n+1}) \left(\, \prod_{i=1}^n p_{\alpha_{i+1} \alpha_i}(t_i) \, \right) p_{\alpha_1 0} + p_{ex,0}(t) \, \delta(x)  \; ,  \nonumber
\label{eq:abs_prob_price}
\end{eqnarray}
while an analogous expression holds for $P^{ex}_n(T,n)$. In particular, the above equation can be recovered from Eq.~(\ref{eq:free_prob_price}) by exploiting the different terminal condition 
\begin{equation}
  \label{eq:4}
 \int_0^{t} dt^\prime (1-p_{\uparrow \alpha}(t^\prime) -p_{\downarrow \alpha}(t^\prime)) \to p_{ex,\alpha}(t) \;,
\end{equation}
which in Laplace space reads
\begin{equation}
  \label{eq:free_abs_subs}
  \frac 1 \omega [1-\hat p_{\uparrow \alpha}(\omega) - \hat p_{\downarrow \alpha}(\omega)] \to \hat p_{ex,\alpha}(\omega) \; .
\end{equation}
The generating functions for this modified problem can then be found as shown in the previous section by taking into account the substitution~(\ref{eq:free_abs_subs}).

We are interested in particular in calculating the statistics of the price and the hitting number at the stopping time $T$, which can recovered by differentiation with respect to $s$ of the generating function evaluated at the point $\omega=0$. We notice in particular that by substituting $\hat p_{\alpha,\alpha^\prime} \to \bar p_{\alpha,\alpha^\prime}$, and by exploiting the relation $\bar p_{ex,\alpha} =1-\bar p_{\uparrow,\alpha} -\bar p_{\downarrow,\alpha} $ in  Eqs.~(\ref{eq:free_av_price}),~(\ref{eq:free_sq_price}),~(\ref{eq:free_av_hit}) and~(\ref{eq:free_sq_hit}), it is possible to use these same formulae even for the execution problem by simply multiplying by $\omega$, thus removing the overall $\omega^{-1}$ factor.

Finally, we would like to obtain the statistics of the stopping time $T$ itself, which can be obtained from the $\omega=0$ value of the generating functions $\Psi(\omega)=\Psi_x(\omega,0)=\Psi_n(\omega,0)$. It reads
\begin{equation}
  \label{eq:abs_gf_time}
  \Psi(\omega) = \hez + \frac 
    { \huz
       (\hdd\heu - \hdu\hed - \heu) +
      \hdz
       (\huu\hed - \hud\heu - \hed)
    }
    {(1-\hdd-\huu-\hud\hdu+\huu \hdd )} \; .
\end{equation}
In the hypothesis in which those functions are at least $q$ times differentiable at $\omega=0$ (valid for $\mu> 0$) one can write
\begin{eqnarray}
  \partial_\omega^q \hat p_{\alpha,\alpha^\prime} |_{\omega=0} &=&  (-1)^q \bar p_{\alpha,\alpha^\prime} \langle t_{\alpha,\alpha^\prime}^q \rangle \; .
\end{eqnarray}
The unconditional hitting time $t_{hit,\alpha^\prime}$ is defined as the time required for any of the boundaries to be hit. Its moment of order $q$ is related to the conditional ones through the relation
\begin{equation}
  \langle t_{hit,\alpha^\prime}^q \rangle = \sum_{\alpha}\bar p_{\alpha,\alpha^\prime} \langle t_{\alpha,\alpha^\prime}^q\rangle \;.
\end{equation}
By employing these definitions, one can prove that
\begin{equation}
  \label{eq:abs_av_time}
  \langle T \rangle =  \frac 1 B \left(  A_\downarrow \langle t_{hit,\downarrow} \rangle + A_\uparrow \langle t_{hit,\uparrow} \rangle + B  \langle t_{hit,0} \rangle \right) \; ,
\end{equation}
and
\begin{eqnarray}
  \label{eq:abs_var_time}
  \langle T^2 \rangle - \langle T \rangle^2 &=& \frac{1}{B} \left( A_\downarrow [\langle t_{hit,\downarrow}^2\rangle - \langle t_{hit,\downarrow}\rangle^2]   + \frac B 2 [ \langle t_{hit,0}^2\rangle - \langle t_{hit,0}\rangle^2] \right)  \\
&+& \frac{1}{B^2} \left(- \frac{(A_\downarrow \langle t_{hit,\downarrow} \rangle + A_\uparrow \langle t_{hit,\uparrow}\rangle )^2}{2}
+ A_\downarrow B (\langle t_{hit,\downarrow} \rangle -\langle t_{hit,0} \rangle )^2 - A_\downarrow B \langle t_{hit,0}\rangle^2 \right) \nonumber \\
&+&  \frac{2}{B^2} \bigg( 
A_\downarrow \bud \langle t_{\uparrow\downarrow}\rangle +
A_\uparrow \buu \langle t_{\uparrow\uparrow}\rangle +
B \huz \langle t_{\uparrow0}\rangle
\bigg) \bigg( (1 - \bdd) \langle t_{hit,\uparrow}\rangle + \bdu \langle t_{hit,\downarrow}\rangle  \bigg) \nonumber \\
&+& \textrm{(} \uparrow \textrm{ exch. } \downarrow \textrm{)} \nonumber \; ,
\end{eqnarray}
where we have defined
\begin{eqnarray}
  \label{eq:aux_var}
  A_\alpha &=&  \bar p_{-\alpha} \bar p_{\alpha,-\alpha} + \bar p_{\alpha} - \bar p_{\alpha} \bar p_{-\alpha,-\alpha} \\
  B &=& 1-\bdd-\buu-\bdu\bud+\buu\bdd \;,
\end{eqnarray}
and introduced the convention $\pm\,(\uparrow)\,=\,\mp\,(\downarrow)$.

\bibliography{apparent_impact}{}
\bibliographystyle{unsrt}

\end{document}